\documentclass[10pt, final, journal, letterpaper, oneside, twocolumn]{IEEEtran}
\usepackage{cite}
\usepackage{amsmath,amssymb,amsfonts}
\usepackage{algorithmic}
\usepackage{graphicx}
\usepackage{textcomp}
\usepackage{float}
\usepackage{subfloat}
\usepackage{subfig}
\usepackage{caption}
\usepackage{epstopdf}
\usepackage{xcolor}
\usepackage{hyperref}
\usepackage{color,soul}

\ifCLASSINFOpdf
\else
\fi
\begin{document}
\title{Experimental Evaluation of Multi-operator RIS-assisted Links in Indoor Environment }

\author{\IEEEauthorblockN{Mir~Lodro\IEEEauthorrefmark{1},~\IEEEmembership{Member,~IEEE,} Jean Baptiste~Gros\IEEEauthorrefmark{2}, Steve~Greedy\IEEEauthorrefmark{1}, Geoffroy~Lerosey\IEEEauthorrefmark{2},
Anas~Al~Rawi\IEEEauthorrefmark{3},
Gabriele~Gradoni\IEEEauthorrefmark{1},~\IEEEmembership{Member,~IEEE}}\\\
\IEEEauthorblockA{\IEEEauthorrefmark{1}{University of Nottingham, NG7 2RD,  United Kingdom}}\\\
\IEEEauthorblockA{\IEEEauthorrefmark{2}{Greenerwave, 75005 Paris, France}}\\\
\IEEEauthorblockA{\IEEEauthorrefmark{3}{OFCOM, London, SE1 9HA, United Kingdom}}

% <-this % stops a space
\thanks{Mir Lodro, Steve Greedy and Gabriele Gradoni are with George Green Institute for Electromagnetics Research-GGIEMR, the University of Nottingham, UK. Jean Baptiste Gros and Geoffroy Lerosey are with Greenerwave, Paris}}% <-this % stops a space

\maketitle

\begin{abstract}
In this work, we present reconfigurable intelligent surface (RIS)-assisted optimization of the multiple links in the same indoor environment. Multiple RISs from different operators can co-exists and handle independent robust communication links in the same indoor environment. We investigated the key performance metrics with the help of two simultaneously operating RIS-empowered robust communication links at different center frequencies in the same indoor environment. We found with the help of bit error rate (BER) and error vector magnitude (EVM) measurements that two operators can co-exist in the same RF environment without seriously impacting  quality of service of users.
\end{abstract}


% \fi
\IEEEpeerreviewmaketitle

\section{Introduction}
\label{secI}
\IEEEPARstart{R}{econfigurable} intelligent surface is a promising future technology that consists of a large number of sub-wavelength and low-cost passive elements whose phase and amplitude can be optimized to increase signal coverage, enhance spectral-efficiency \cite{you2020energy}, perform accurate RF sensing \cite{hu2020reconfigurable} and increase PHY layer security \cite{chen2019intelligent}. Each element in the RIS can introduce phase shift to impinging electromagnetic waves and the collective phase shifts introduced by all the elements can perform passive beamforming to the desired user. Coverage extension \cite{popov2021experimental} and localization are the two important benefits of RIS that shall pave the way for sixth generation (6G) wireless communication \cite{strinati2021wireless}. In order to demonstrate RIS-assisted communication system different experimental setup and  prototypes have been created at sub-6GHz and at mmWaves. For example, authors in \cite{dai2020reconfigurable} have proposed a 256 2-bit element RIS and presented experimental results at 2.3 GHz and 28.5 GHz. They have performed RIS-assisted measurements using single Tx-Rx link by transmitting OFDM modulated waveforms over-the-air in an indoor environment. Several other works have provided their experimental setups designed for a single-carrier frequency at sub-6 GHz and at mmWaves \cite{gros2021reconfigurable},\cite{alexandropoulos2021reconfigurable}\cite{fara2021reconfigurable}\cite{pei2021ris}\cite{amri2021reconfigurable}. Authors in \cite{pei2021ris} have performed single-carrier RIS-assisted over-the-air test in both indoor and outdoor environment at operating frequency of 5.8 GHz. Similarly, authors in \cite{amri2021reconfigurable} have performed RIS-assisted measurements in an indoor environment and performed KPI optimization. Recent experimental research on RIS-assisted communication systems remains focused on passive beamforming and KPI optimization of single Tx-Rx links. However, the question of understanding the interference control among multiple RIS-assisted links in the same environment remains outstanding. We use an experimental approach to address the issue of interference among co-existing multiple operator communication links in the same indoor environment. In the same RF environment both the RISs from two operators are exposed to transmitted EM waves from both transmitters at the same time. In this setup where an independent phase optimization of one link may degrade the quality of another communication link. It's highly important to experimentally understand the effect of multiple RIS-assisted links in the same RF environment.

We have divided our work into six sections. After the introduction in section \ref{secI}, we have discussed indoor RF environment in section \ref{secII}, where RIS-assisted measurements are conducted. Section \ref{secIII} discusses communication system KPIs that have been optimized. Section \ref{secIV} gives detailed discussion of the experimental setup and the measurement procedure. Section \ref{secV} is dedicated to results and discussion section where the details about the two configurations of RIS are presented. Section \ref{secVI} is about the lessons learned from the measurement campaign, the conclusion and future aspects of the research.

\section{Indoor Environment}
\label{secII}
\begin{figure}
    \centering
    \includegraphics[width=0.9\columnwidth]{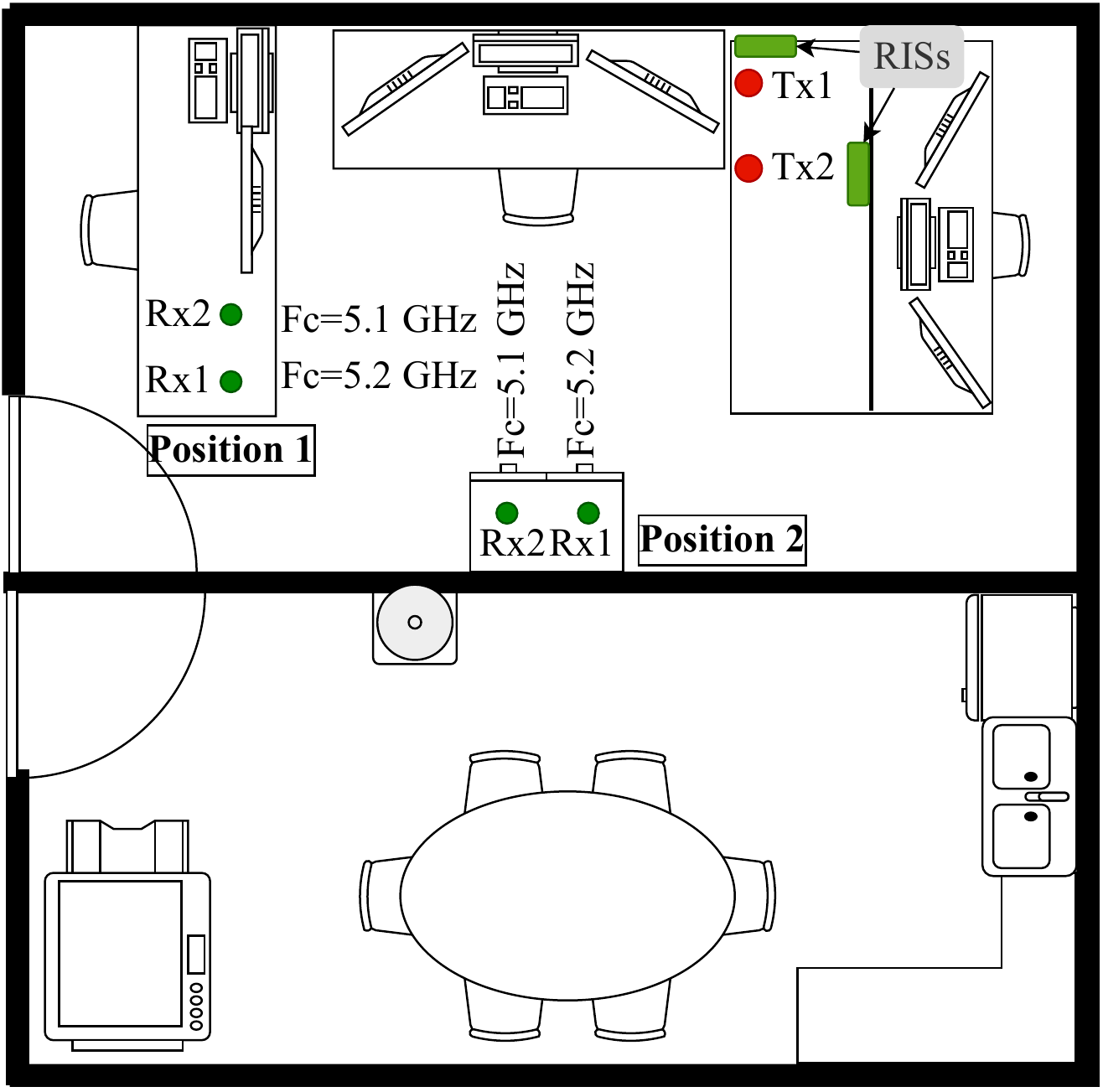}
    \caption{Indoor environment floor plan.}
    \label{fig:floorplan}
\end{figure}

The RIS-empowered two link optimization was performed in an indoor office environment. The floor plan of the RF environment is shown in Fig. \ref{fig:floorplan}. The indoor environment is an office environment in a multi-story building and the measurement setup is created in an office room. The office room consists of wooden furniture including desks, chairs, small cabinets and IT equipment such as monitors and the host PCs. The indoor environment for sub-6GHz is a rich-scattering environment where the two Tx-Rx links are assisted using RISs. Each communication link is assisted independently by a separate RIS. There exist direct link between transmitter and receiver in addition to cascaded channel Tx-RIS-Rx. The direct link is weak enough to be considered negligible and hence the KPI optimization is performed only using indirect i.e. Tx-RIS-Rx channel. 
\section{EVM and BER Measurement}
\label{secIII}
EVM and BER are two key performance indicators (KPIs) in digital wireless communication systems allowing greater understanding of modulation quality and the end-to-end communication system performance metric. In baseband model, the EVM can be measured at different points in the digital receiver before frame synchronization and data recovery stage. The modulation techniques with constant magnitude such as QPSK, the EVMs are normalized to constellation maximum and those with multiple magnitude levels the EVM is normalized to RMS level of I/Q reference symbols \cite{chen2022calibrated}.
Bit error rate (BER) is an end-to-end communication system performance metric. The BER is measured after all the RF impairments have been corrected. Additionally, the BER is measured after the frame equalization, phase ambiguity and data recovery steps have been correctly applied. BER is an ultimate performance indicator which guarantees all the transmitter steps have been reversed at the receiver including RF impairment compensation.
\begin{equation}
    BER=\frac{\textrm{number of received bits in error}}{\textrm{total number of transmitted bits}}
\end{equation}

\section{Experimental Setup}
\label{secIV}
\begin{figure}
    \centering
    \includegraphics[width=0.9\columnwidth]{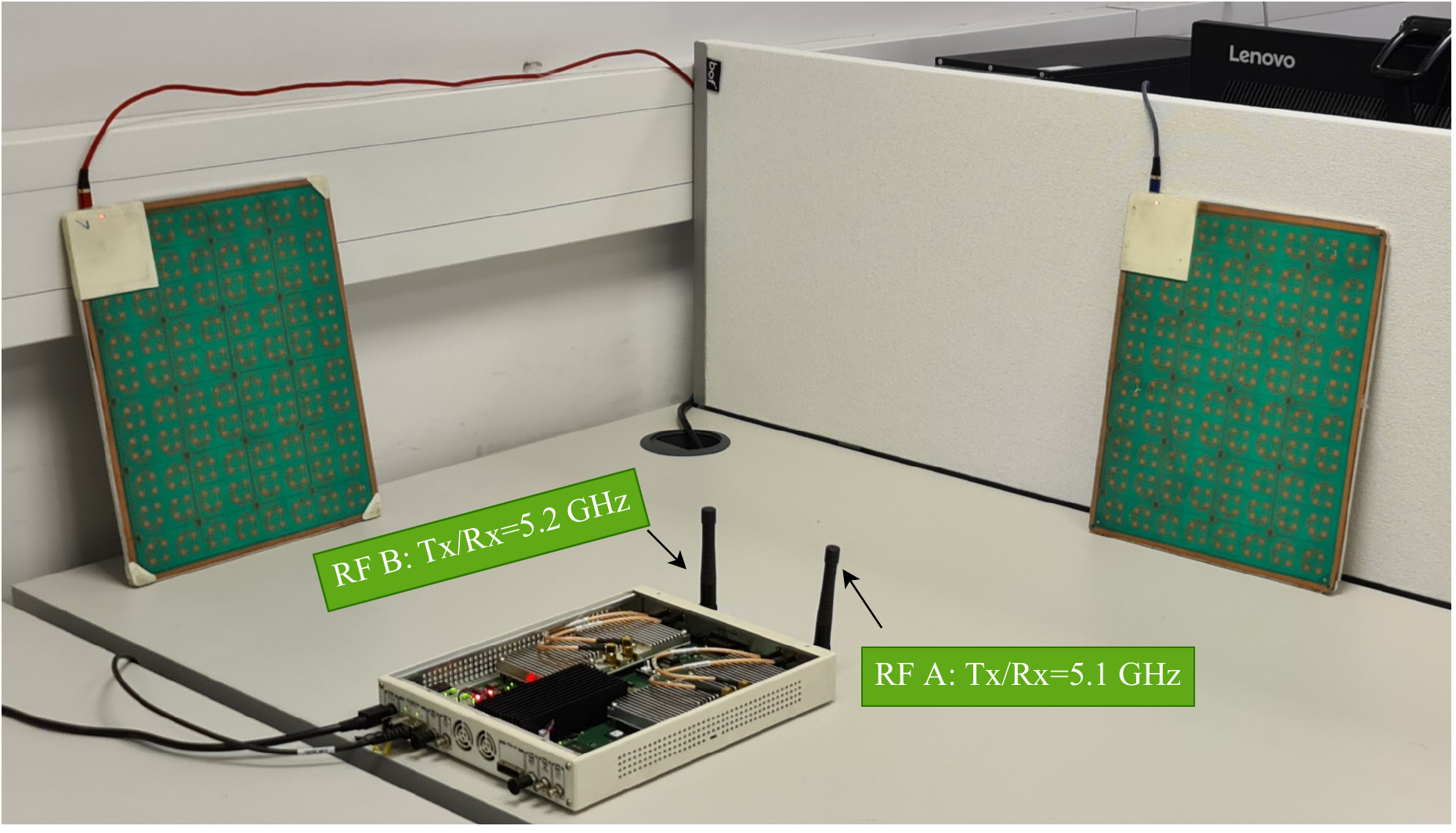}
    \caption{Photograph of the experimental setup of the two separated RISs in relation to Tx1 and Tx2.}
    \label{fig:exp}
\end{figure}

In this work, we used two full-duplex X310 USRPs from Ettus Research. Each of the X310 USRP was equipped with two UBX-160 RF cards that can support 2x2 MIMO links. We used separate USRPs for the transmitter and the receiver in order to avoid inter-board cross-talk. The two RISs and the Tx USRPs and the Rx USRPs were connected to the host PC. We transmitted QPSK modulated waveforms from each channel of the Tx USRP independently at different center frequencies in the indoor environment. The two links operate at adjacent center frequencies and are a single-input single-output (SISO) links. The complex samples were captured by another Rx USRP whose Rx channels were tuned to the corresponding channels of the Tx USRPs. The transmitted waveforms and the received waveforms are generated and processed in a separate MATLAB session. The complex samples are transferred to and from the host PC at a sample rate of 400 kS/s. Hence, the maximum data-transfer rate is 12.8 Mbps for the sample size of 16-bit for I and Q samples. The captured QPSK waveforms are down-converted and processed in the baseband model where the baseband waveforms are compensated for RF impairments such as coarse frequency compensation, fine-frequency compensation and the timing synchronization. The measured waveforms and the EVMs of two links are recorded after all the RF impairments have been rectified. We recorded our previous experience with current baseband model in \cite{lodro2020near}. For customizing the radio channel we used two 5.2 GHz 1-bit RISs patented by Greenerwave, Paris that resonates at 5.2 GHz. The RISs have been used in previous experiments on link-optimization and changing the boundary conditions in a chaotic metal-enclosure exhibiting rich-scattering \cite{lodro2021reconfigurable}\cite{gros2020tuning}. 
Fig. \ref{fig:exp} shows experimental setup of two separated RIS-assisted links measurement setup. Each metasurface consists of 76 binary pixels whose reflection co-efficient can be changed by controlling the biase voltage of PIN diodes. Binary phase shift is applied to RIS elements using RIS controller. The RIS controller connected to the host PC using USB2.0 cable and it receives PIN diode mapping instructions from a MATLAB program. The two RIS-assisted links use monopoles for the Tx and Rx links such that the two RISs are exposed to transmitted signals from the cross-links.
\section{Discussion and Results}
\label{secV}
We conducted EVM and BER optimization by optimizing the configurations of two RISs where each RIS supporting separate links at designated center frequency. The RISs are designed to operate at center frequency of 5.2 GHz. Multi-operator link optimization is performed by optimizing configurations of each RIS supporting independent link at adjacent frequencies. Two scenarios at different center frequencies are investigated. Co-located RISs when the two RISs are placed side by side and they are at the same distance from the transmitters. In separated RISs scenario the two RISs are placed along the axes that are orthogonal to each other. 

\subsection{Co-located RISs}
\begin{figure}
    \centering
    \subfloat[]{\label{fig:ini_const}\includegraphics[width=0.5\columnwidth]{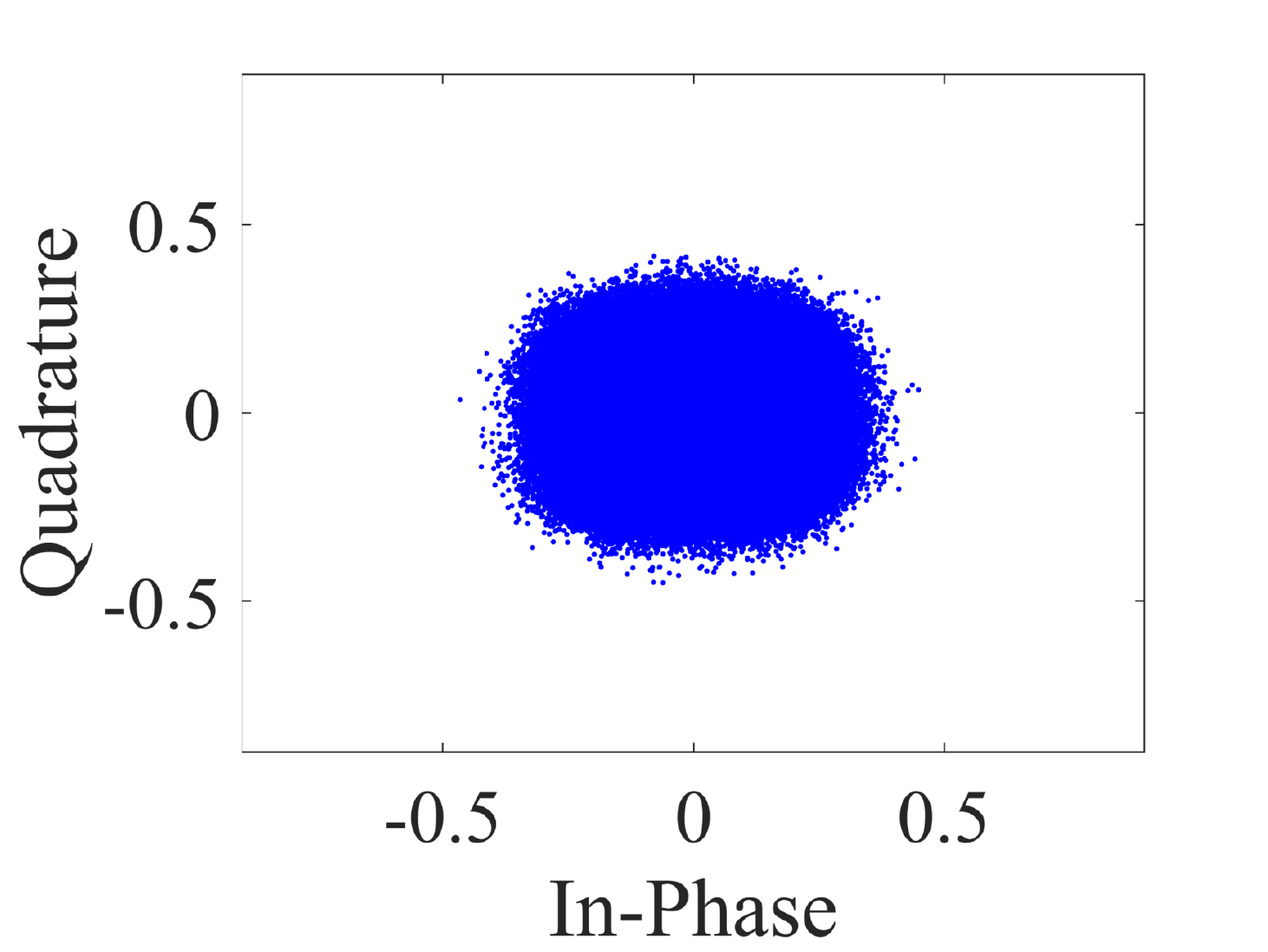}}
    \subfloat[]{\label{fig:opt_const}\includegraphics[width=0.5\columnwidth]{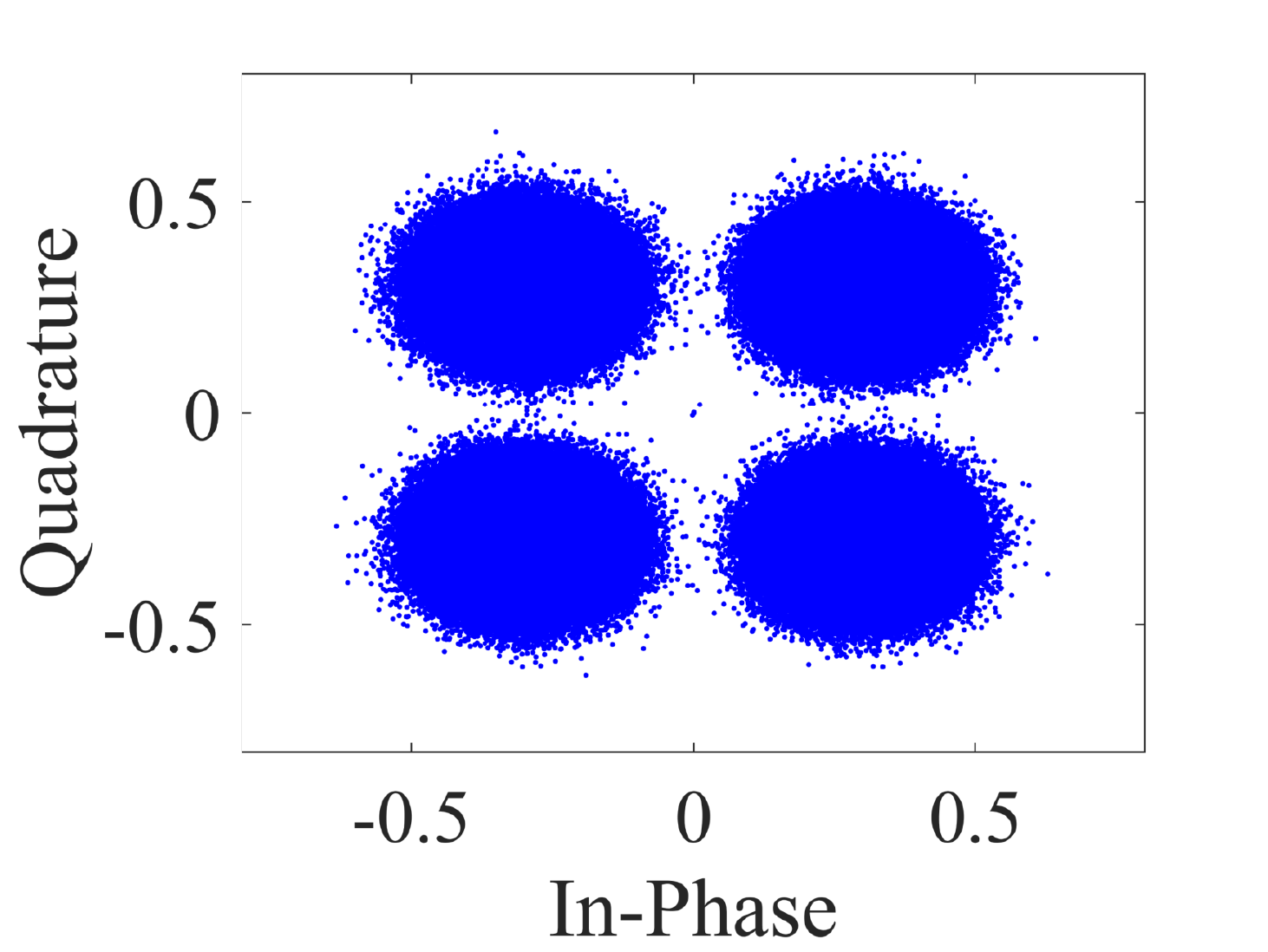}}\\
    \subfloat[]{\label{fig:ini_const52}\includegraphics[width=0.5\columnwidth]{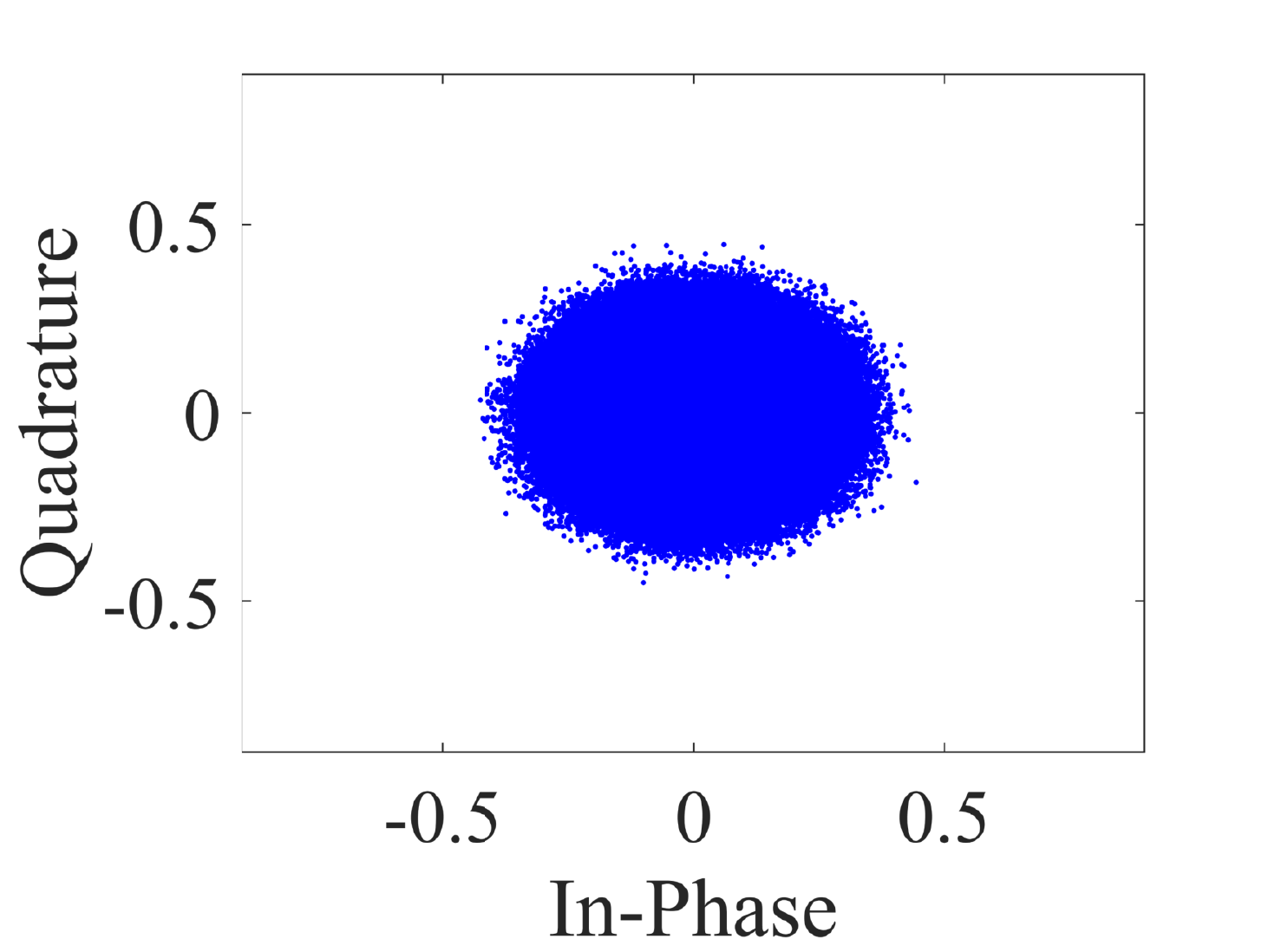}}
    \subfloat[]{\label{fig:opt_const52}\includegraphics[width=0.5\columnwidth]{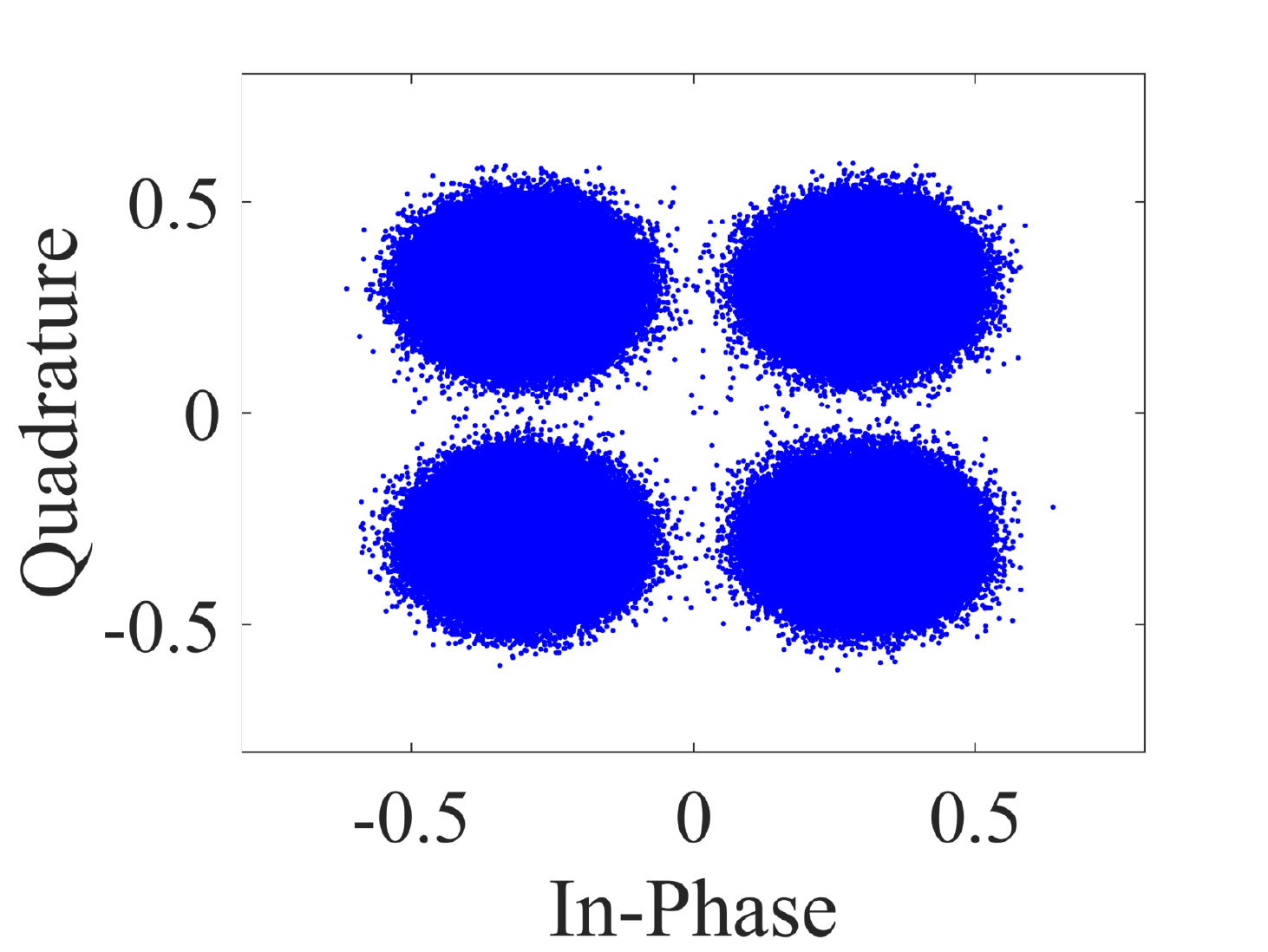}}
    \caption{Typical QPSK constellations (a) typical initial QPSK constellation at 5.1 GHz (b) typical optimized QPSK constellation at 5.1 GHz (c) typical initial QPSK constellation at 5.2 GHz (d) typical optimized QPSK constellation at 5.2 GHz.}
    \label{fig:qpsk_const}
\end{figure}

We considered two co-located RISs supporting links at the center frequencies of 5.1 GHz and 5.2 GHz respectively. In co-located scenario we placed two RIS side by side and established two independent communication links. The Tx side of both the links were placed closely to co-located RIS. The receivers of both the links were also co-located and the receivers were placed at a distance of 4 meters from the RISs. Each RISs assisted an independent link operating at different carrier frequencies with a minimum gap of 50 MHz. For co-located scenario, we performed KPI optimization at 5.1 GHz and 5.2 GHz respectively. We started with poor channel condition for both the links. It was visually observed from the constellation diagrams at the starting position the received QPSK symbols were centered around zero on the IQ plane as shown in Fig.\ref{fig:ini_const}. After RIS optimization the QPSK symbols were received in their corresponding quadrant on the IQ plane (see Fig. \ref{fig:opt_const}).

\begin{figure}
    \centering
    \subfloat[]{\label{fig:co_inst_EVM}\includegraphics[width=0.5\columnwidth]{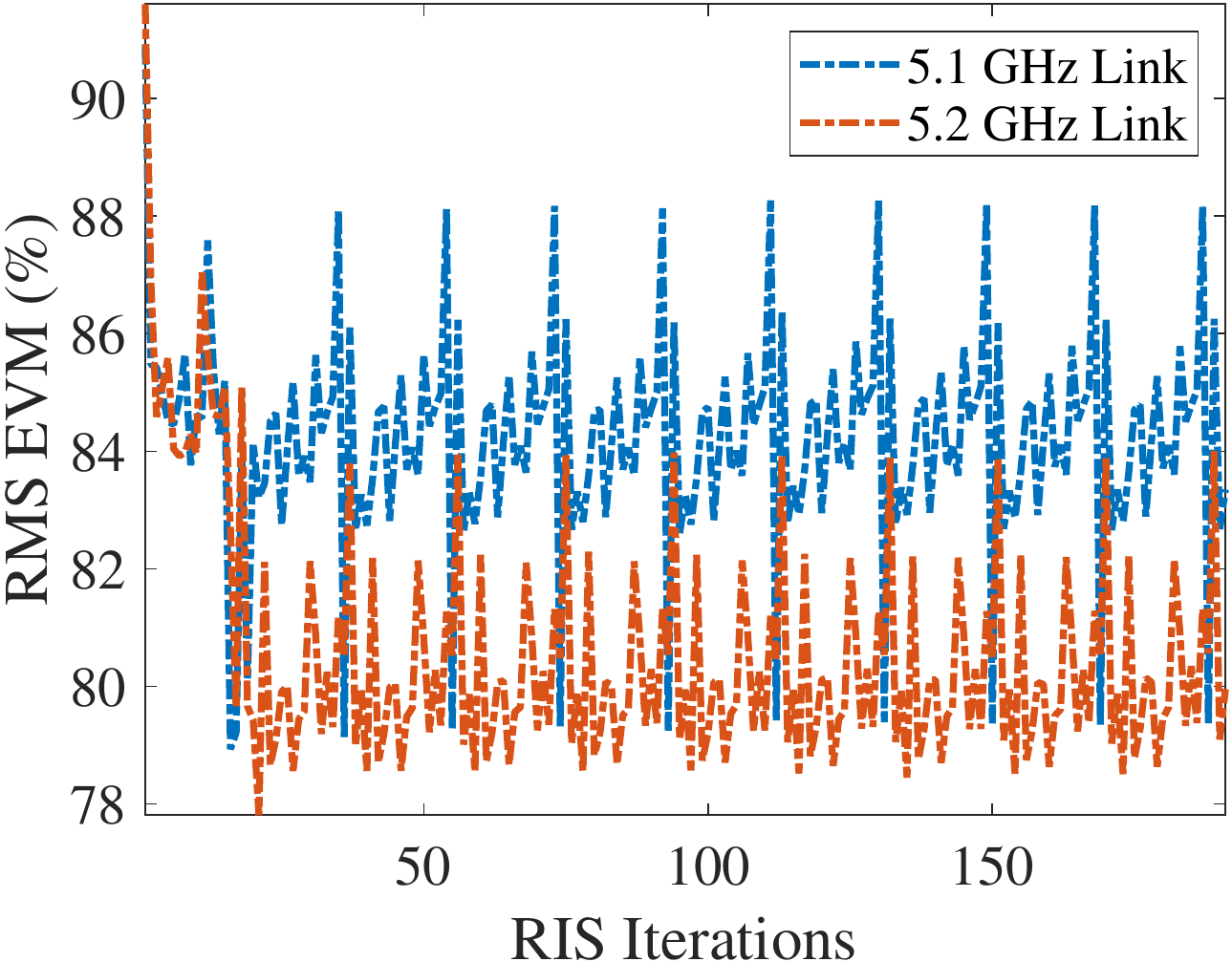}}
    \subfloat[]{\includegraphics[width=0.5\columnwidth]{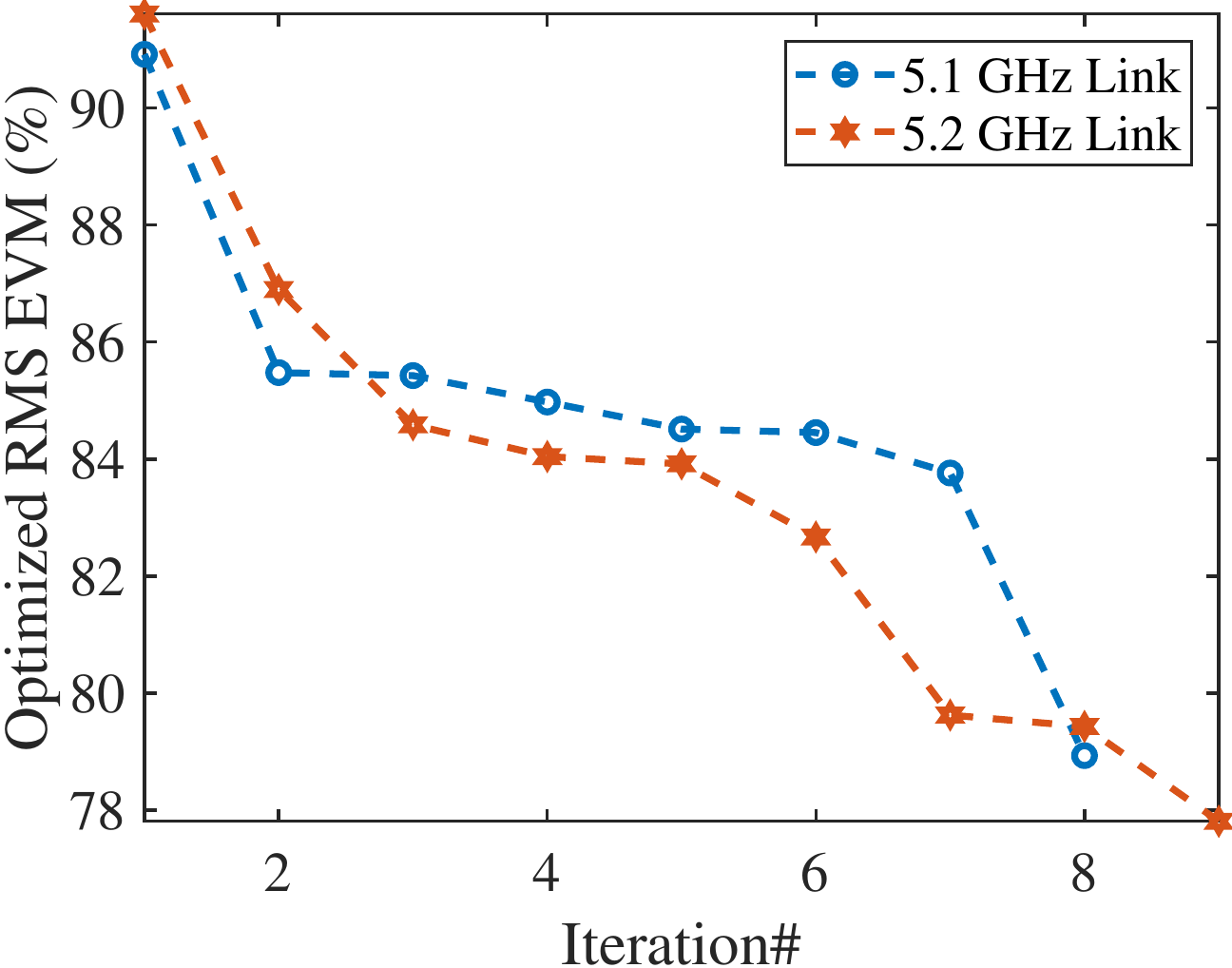}}
    \caption{Co-located RISs scenario: QPSK EVM at 5.1 GHz and 5.2 GHz respectively. The center frequency difference is 100 MHz (a) instantaneous EVM (b) optimized EVM.}
    \label{fig:co_evm}
\end{figure}

\begin{figure}
    \centering
    \subfloat[]{\label{fig:5a}\includegraphics[width=0.5\columnwidth]{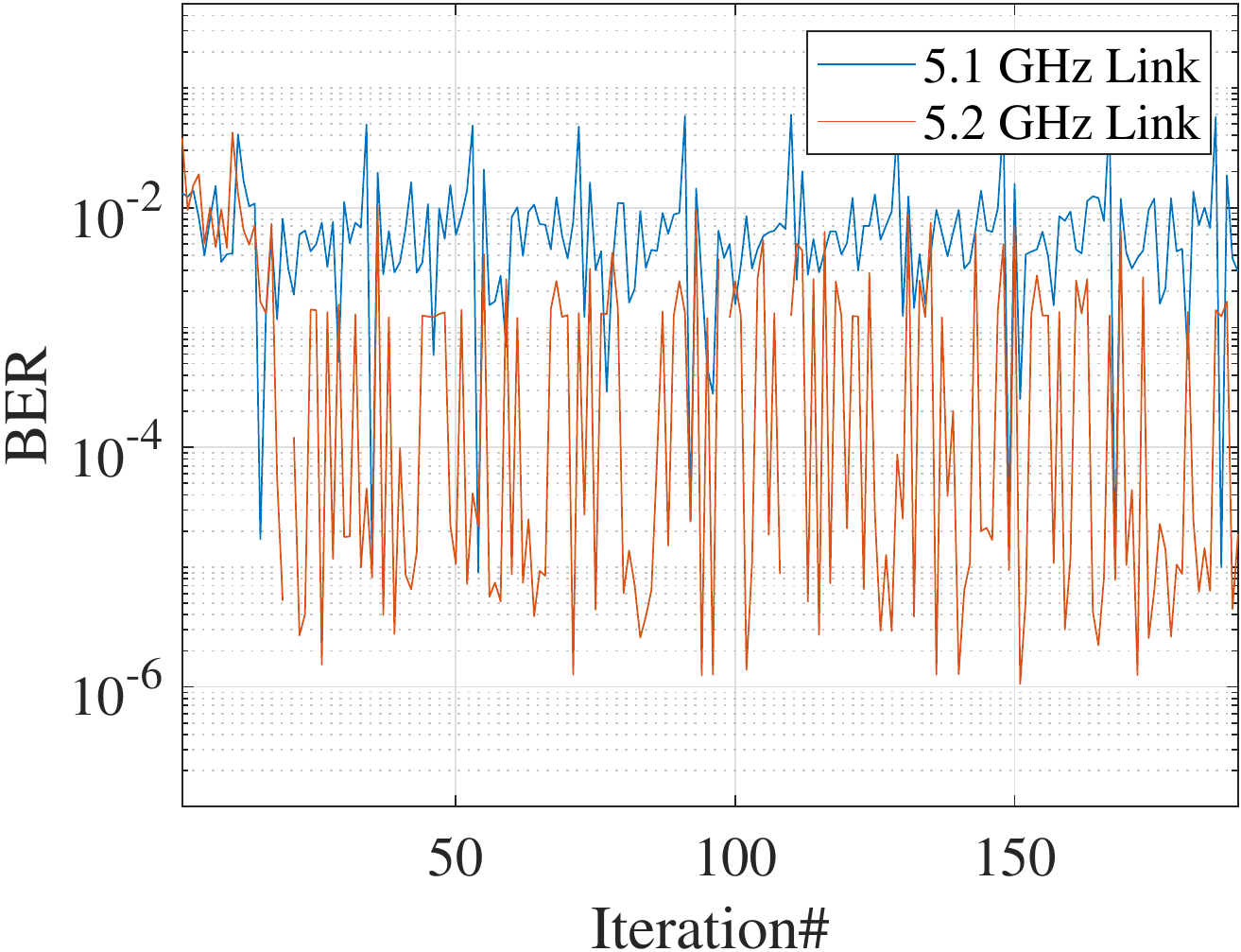}}
    \subfloat[]{\includegraphics[width=0.5\columnwidth]{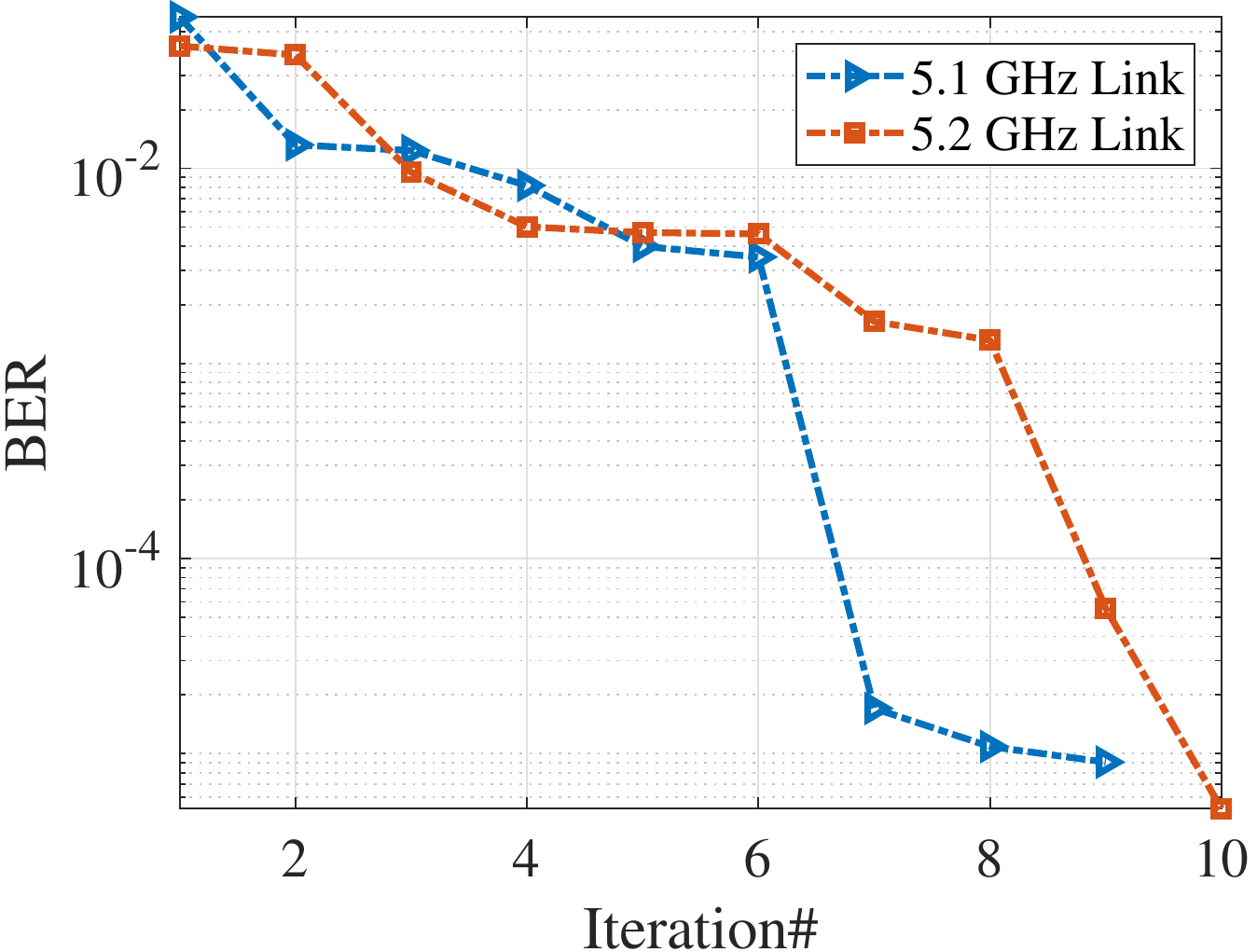}}
    \caption{Co-located RISs scenario: QPSK BER at 5.1 GHz and 5.2 GHz respectively. The center frequency difference is 100 MHz (a) instantaneous BER (b) optimized BER.}
    \label{fig:co_ber}
\end{figure}

Fig. \ref{fig:co_evm} shows the instantaneous RMS EVM and the optimized RMS EVM at 5.1 GHz and 5.2 GHz links respectively in a co-located RISs scenario. Instantaneous EVM of the two links decreases and the RMS EVM for 5.2 GHz link is consistently less than the 5.1 GHz link. Fig. \ref{fig:co_ber} shows instantaneous BER and the optimized BER of two RIS-assisted links in the co-located RISs scenario. Fig. \ref{fig:5a} shows the instantaneous BER of 5.2 GHz link is consistently lower than the 5.1 GHz link because of the RIS peak response.

\subsection{Separated RISs}
\begin{figure}
    \centering
    \subfloat[]{\label{fig:7a}\includegraphics[width=0.5\columnwidth]{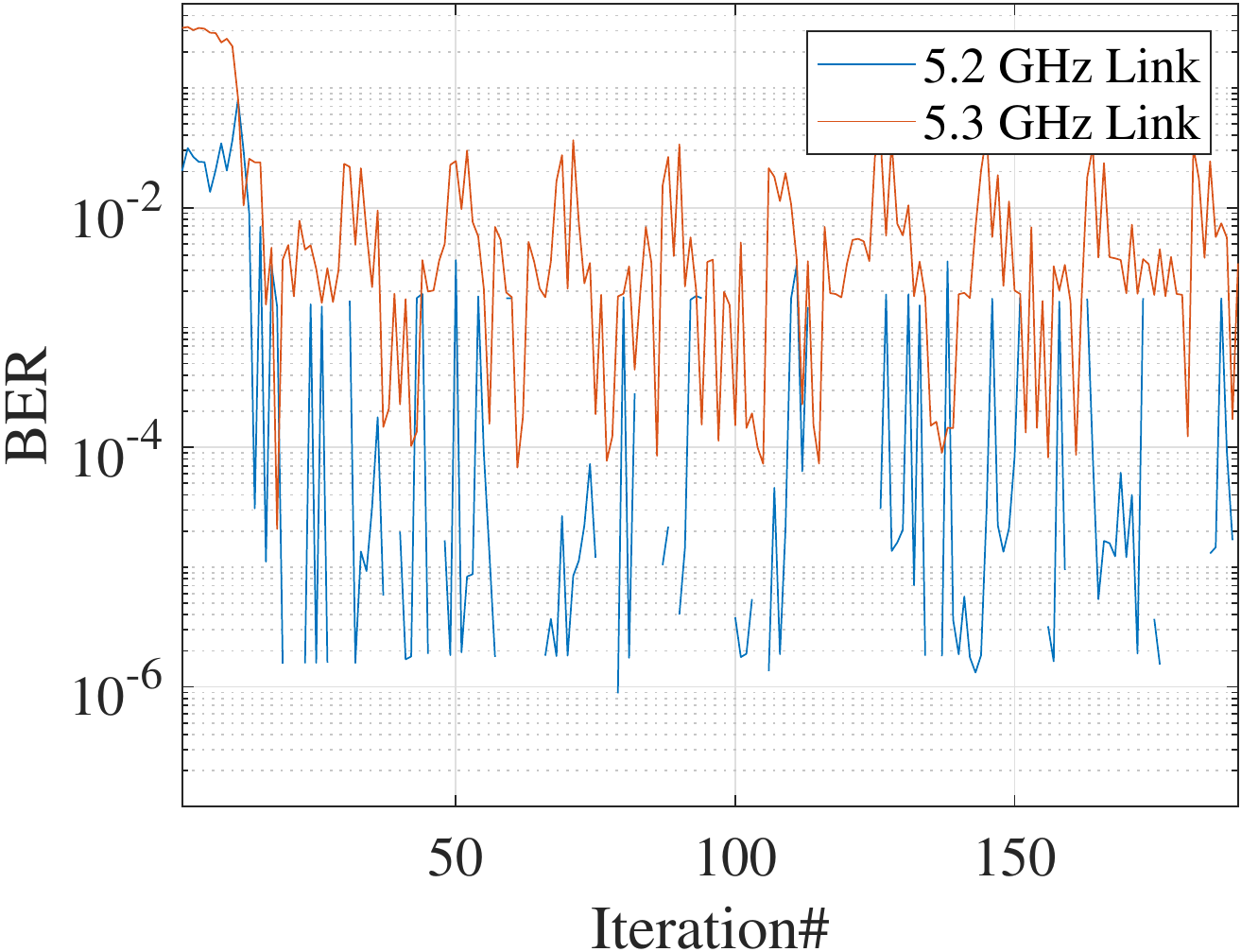}}
    \subfloat[]{\includegraphics[width=0.5\columnwidth]{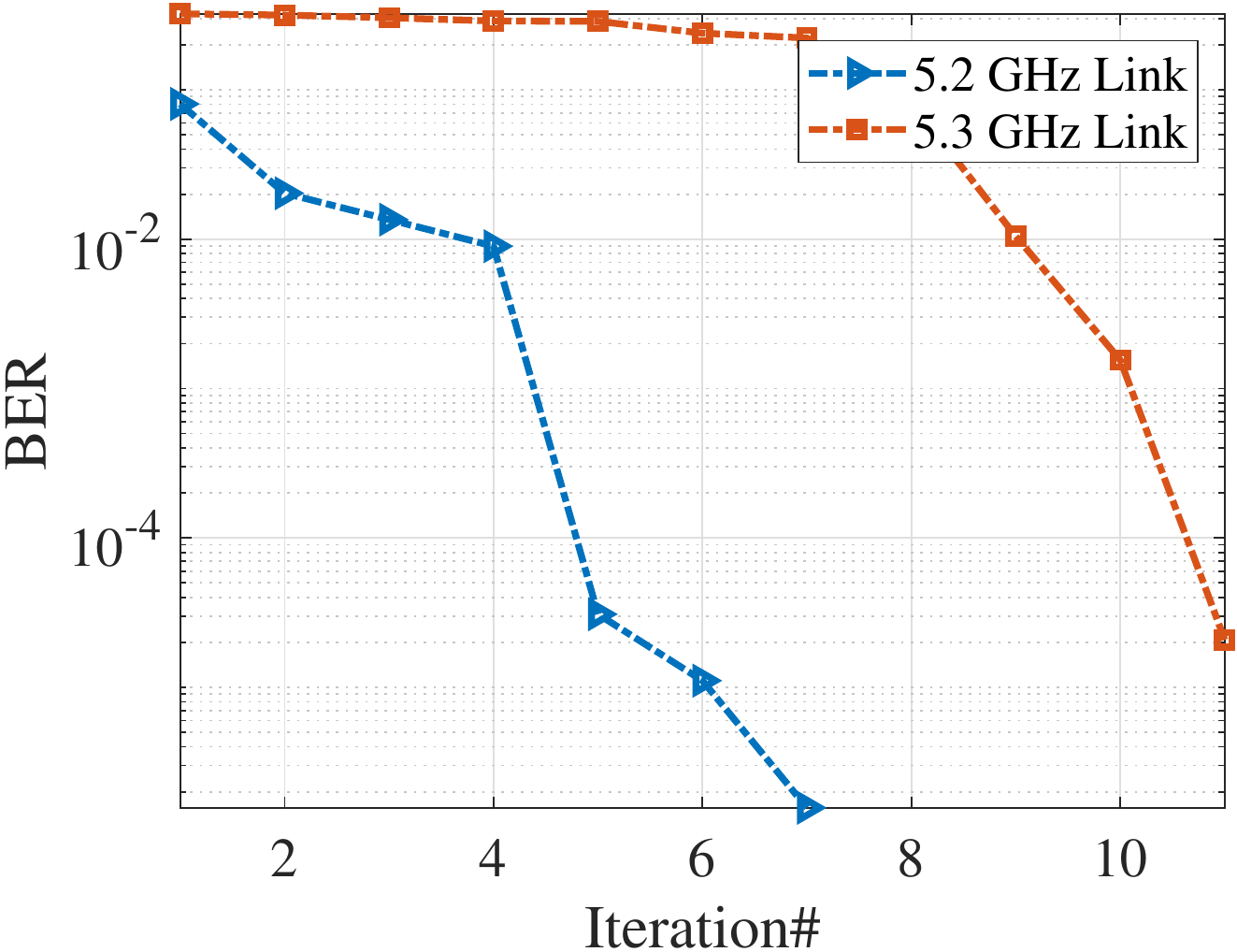}}
    \caption{Separated RISs scenario: QPSK BER at 5.2 GHz and 5.3 GHz respectively. The center frequency difference is 100 MHz (a) instantaneous BER (b) optimized BER.}
    \label{fig:sep_52_53_ber}
\end{figure}

We considered separated RISs placed along x-axis and y-axis respectively. We repeated the measurements for several frequencies centered around RISs operating frequencies. We observed RISs optimization was more effective at frequencies closer to its operating frequencies. For the fixed USRP transmit gains we observed low-edge frequencies with better RF propagation specially when the two links were separately operating with separation of 100 MHz i.e. at 5.1 GHz and 5.2 GHz respectively. We further reduced the gap of two RF frequencies to 50 MHz. The two links were centered around 5.2 GHz and we observed almost same RF effects. We visually observed this from the QPSK constellation diagram. The EVM optimization of two RISs links closely followed each other. Fig. \ref{fig:sep_52_53_ber} shows instantaneous BER and the optimized BER of two RIS-assisted links in the separated RISs scenario. Similar to Fig. \ref{fig:5a}, Fig. \ref{fig:7a} shows the instantaneous BER of 5.2 GHz link is consistently lower than the 5.3 GHz link because of RIS peak response.

\begin{figure}
    \centering
    \subfloat[]{\label{fig:ini_evm_52_53}\includegraphics[width=0.5\columnwidth]{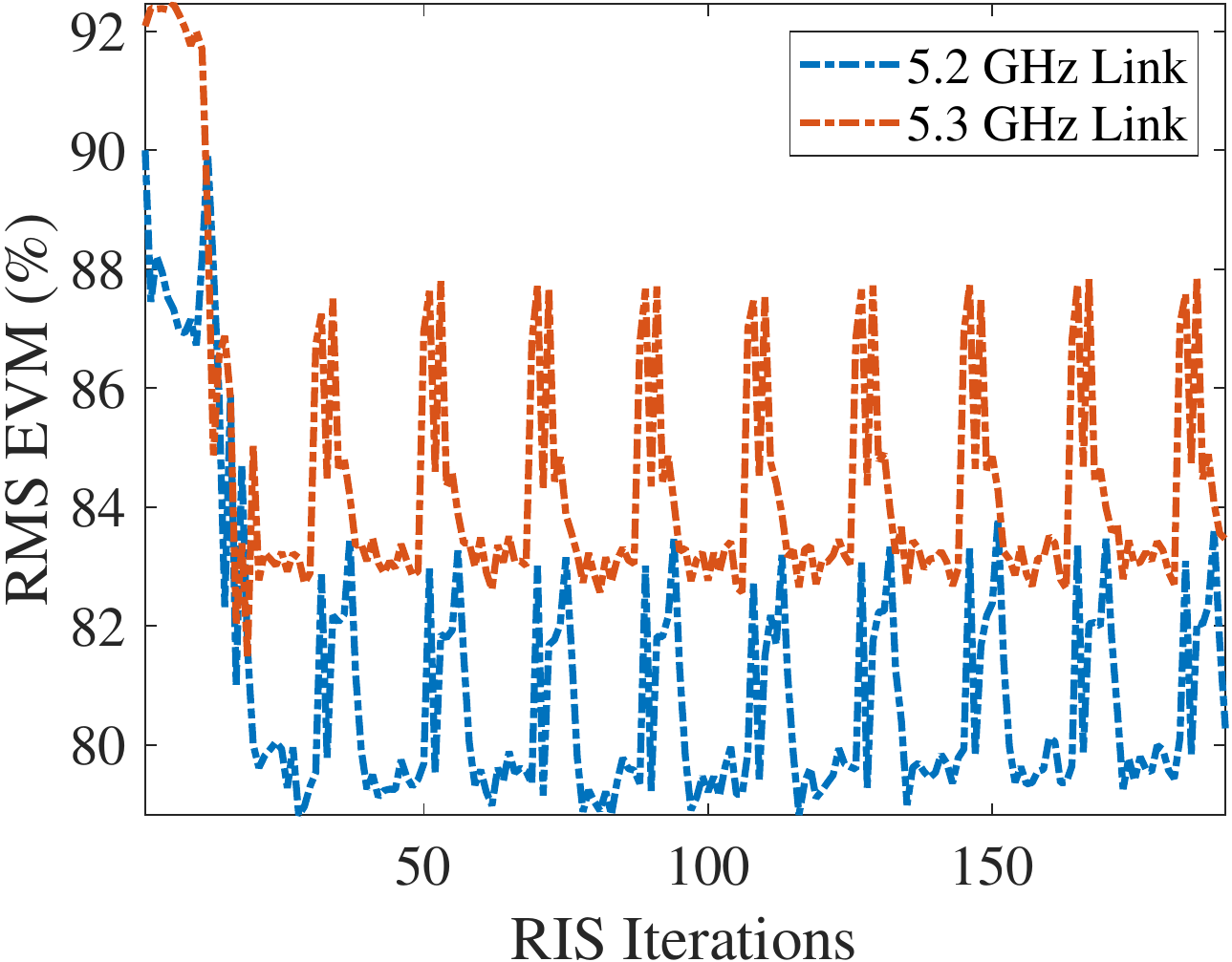}}
    \subfloat[]{\includegraphics[width=0.5\columnwidth]{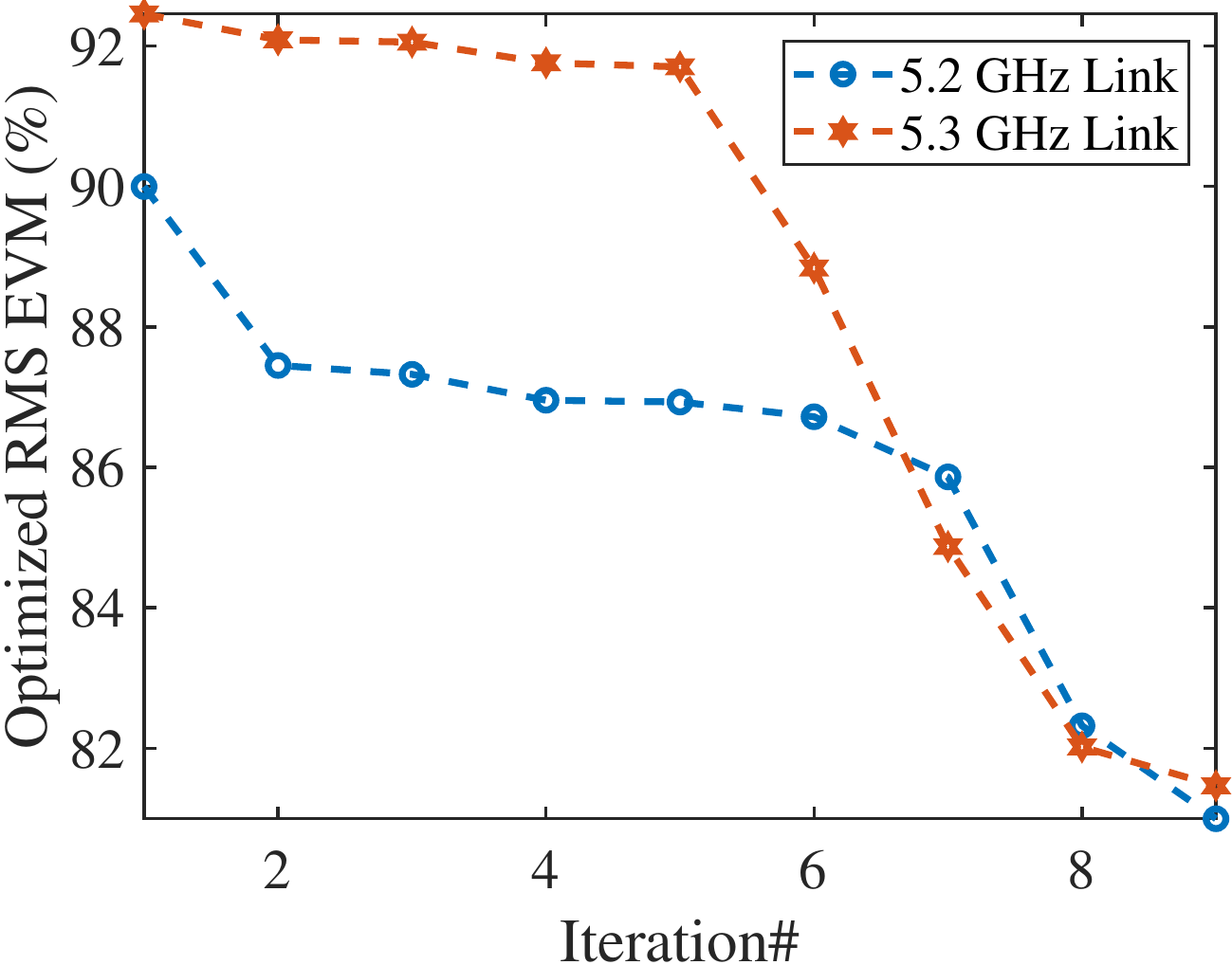}}
    \caption{Separated RISs scenario: QPSK EVM at 5.2 GHz and 5.3 GHz respectively. The center frequency difference is 100 MHz (a) instantaneous EVM (b) optimized EVM. }
    \label{fig:sep_52_53_evm}
\end{figure}

\begin{figure}[!htbp] 
    \centering
    \subfloat[]{\label{fig:ini_evm_5175_5225_50}\includegraphics[width=0.5\columnwidth]{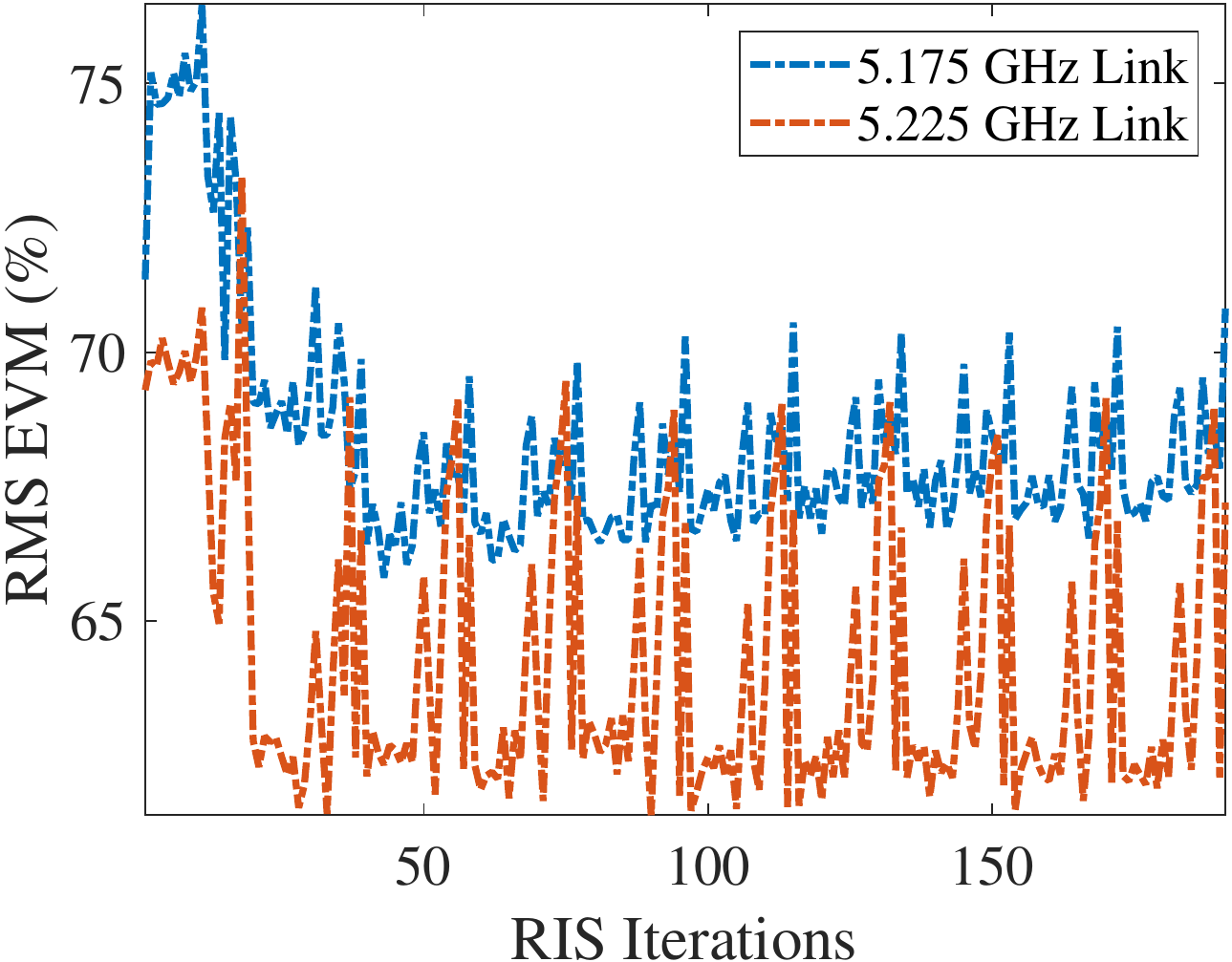}}
    \subfloat[]{\includegraphics[width=0.5\columnwidth]{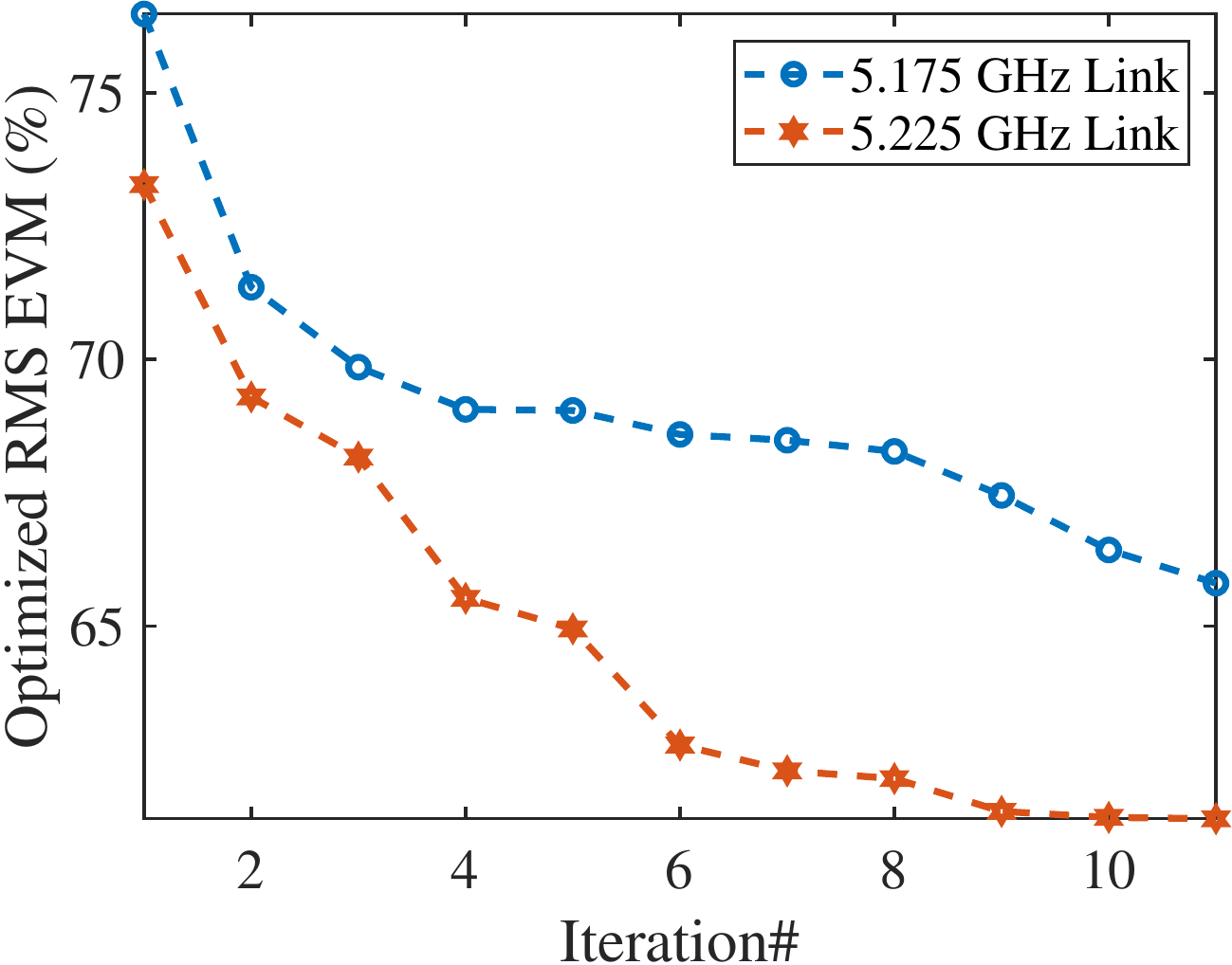}}
    \caption{Separated RISs QPSK EVM at 5.175 GHz and 5.225 GHz respectively. The center frequency difference is 50 MHz (a) instantaneous EVM (b) optimized EVM.}
    \label{fig:sep_5175_5225_evm}
\end{figure}

We also measured QPSK EVM and BER in separated RISs scenario by changing center frequency of the SDR to 5.175 GHz and 5.225 GHz. The center frequency difference of 50 MHz. The separated RISs scenario optimization of QPSK EVM and QPSK BER was performed. We observed the similar pattern as in previous cases. We observed BER only at initial RISs iterations then no BERs were observed as the link started with favourable propagation but the EVM optimization followed same repetitive patterns. The measured QPSK EVM and the optimized QPSK EVM in separated RIS scenario is shown in Fig.\ref{fig:sep_5175_5225_evm}. Both links have been optimized.
From Fig. \ref{fig:co_inst_EVM}, Fig. \ref{fig:ini_evm_52_53} and Fig. \ref{fig:ini_evm_5175_5225_50} we understand that the RIS effectively optimizes the links at 5.2 GHz or closer to 5.2 GHz. The instantaneous EVM is higher for center frequency centered on or closer to  5.2 GHz independent of whether the two links started with poor or favourable propagation conditions.
\section{Conclusion}
\label{secVI}
An experimental study was carried out to understand multiple RIS-assisted links in the same indoor environment. We conducted RIS-assisted multiple link optimization in different scenarios using SDR based testbed. In all the experiments RIS configurations were optimized to decrease the EVM and the BER of each communication link. From our measurements experience, we understand that multi-operator RIS-assisted links can be sustained in the same environment without affecting or degrading the QoS of served users.

\section*{Acknowledgment}
This work has been supported by the European Commission through the H2020 RISE-6G Project under Grant 101017011.

% \appendices

% % you can choose not to have a title for an appendix
% % if you want by leaving the argument blank
% \section{}

\ifCLASSOPTIONcaptionsoff
  \newpage
\fi



%\bibliographystyle{IEEEtran}
\bibliographystyle{IEEEtran}
\bibliography{two_operators.bib}

\begin{thebibliography}{10}
\providecommand{\url}[1]{#1}
\csname url@samestyle\endcsname
\providecommand{\newblock}{\relax}
\providecommand{\bibinfo}[2]{#2}
\providecommand{\BIBentrySTDinterwordspacing}{\spaceskip=0pt\relax}
\providecommand{\BIBentryALTinterwordstretchfactor}{4}
\providecommand{\BIBentryALTinterwordspacing}{\spaceskip=\fontdimen2\font plus
\BIBentryALTinterwordstretchfactor\fontdimen3\font minus
  \fontdimen4\font\relax}
\providecommand{\BIBforeignlanguage}[2]{{%
\expandafter\ifx\csname l@#1\endcsname\relax
\typeout{** WARNING: IEEEtran.bst: No hyphenation pattern has been}%
\typeout{** loaded for the language `#1'. Using the pattern for}%
\typeout{** the default language instead.}%
\else
\language=\csname l@#1\endcsname
\fi
#2}}
\providecommand{\BIBdecl}{\relax}
\BIBdecl

\bibitem{you2020energy}
L.~You, J.~Xiong, D.~W.~K. Ng, C.~Yuen, W.~Wang, and X.~Gao, ``Energy
  efficiency and spectral efficiency tradeoff in ris-aided multiuser mimo
  uplink transmission,'' \emph{IEEE Transactions on Signal Processing},
  vol.~69, pp. 1407--1421, 2020.

\bibitem{hu2020reconfigurable}
J.~Hu, H.~Zhang, B.~Di, L.~Li, K.~Bian, L.~Song, Y.~Li, Z.~Han, and H.~V. Poor,
  ``Reconfigurable intelligent surface based rf sensing: Design, optimization,
  and implementation,'' \emph{IEEE Journal on Selected Areas in
  Communications}, vol.~38, no.~11, pp. 2700--2716, 2020.

\bibitem{chen2019intelligent}
J.~Chen, Y.-C. Liang, Y.~Pei, and H.~Guo, ``Intelligent reflecting surface: A
  programmable wireless environment for physical layer security,'' \emph{IEEE
  Access}, vol.~7, pp. 82\,599--82\,612, 2019.

\bibitem{popov2021experimental}
V.~Popov, M.~Odit, J.-B. Gros, V.~Lenets, A.~Kumagai, M.~Fink, K.~Enomoto, and
  G.~Lerosey, ``Experimental demonstration of a mmwave passive access point
  extender based on a binary reconfigurable intelligent surface,'' \emph{arXiv
  preprint arXiv:2107.02087}, 2021.

\bibitem{strinati2021wireless}
E.~C. Strinati, G.~C. Alexandropoulos, V.~Sciancalepore, M.~Di~Renzo,
  H.~Wymeersch, D.-T. Phan-Huy, M.~Crozzoli, R.~D'Errico, E.~De~Carvalho,
  P.~Popovski, P.~Di~Lorenzo, L.~Bastianelli, M.~Belouar, J.~E. Mascolo,
  G.~Gradoni, S.~Phang, G.~Lerosey, and B.~Denis, ``Wireless environment as a
  service enabled by reconfigurable intelligent surfaces: The rise-6g
  perspective,'' in \emph{2021 Joint European Conference on Networks and
  Communications \& 6G Summit (EuCNC/6G Summit)}, 2021, pp. 562--567.

\bibitem{dai2020reconfigurable}
L.~Dai, B.~Wang, M.~Wang, X.~Yang, J.~Tan, S.~Bi, S.~Xu, F.~Yang, Z.~Chen,
  M.~Di~Renzo \emph{et~al.}, ``Reconfigurable intelligent surface-based
  wireless communications: Antenna design, prototyping, and experimental
  results,'' \emph{IEEE Access}, vol.~8, pp. 45\,913--45\,923, 2020.

\bibitem{gros2021reconfigurable}
J.-B. Gros, V.~Popov, M.~A. Odit, V.~Lenets, and G.~Lerosey, ``A reconfigurable
  intelligent surface at mmwave based on a binary phase tunable metasurface,''
  \emph{IEEE Open Journal of the Communications Society}, vol.~2, pp.
  1055--1064, 2021.

\bibitem{alexandropoulos2021reconfigurable}
G.~C. Alexandropoulos, N.~Shlezinger, and P.~Del~Hougne, ``Reconfigurable
  intelligent surfaces for rich scattering wireless communications: Recent
  experiments, challenges, and opportunities,'' \emph{IEEE Communications
  Magazine}, vol.~59, no.~6, pp. 28--34, 2021.

\bibitem{fara2021reconfigurable}
R.~Fara, D.-T. Phan-Huy, P.~Ratajczak, A.~Ourir, M.~Di~Renzo, and J.~De~Rosny,
  ``Reconfigurable intelligent surface-assisted ambient backscatter
  communications--experimental assessment,'' in \emph{2021 IEEE International
  Conference on Communications Workshops (ICC Workshops)}.\hskip 1em plus 0.5em
  minus 0.4em\relax IEEE, 2021, pp. 1--7.

\bibitem{pei2021ris}
X.~Pei, H.~Yin, L.~Tan, L.~Cao, Z.~Li, K.~Wang, K.~Zhang, and E.~Bj{\"o}rnson,
  ``Ris-aided wireless communications: Prototyping, adaptive beamforming, and
  indoor/outdoor field trials,'' \emph{IEEE Transactions on Communications},
  vol.~69, no.~12, pp. 8627--8640, 2021.

\bibitem{amri2021reconfigurable}
M.~M. Amri, N.~M. Tran, and K.~W. Choi, ``Reconfigurable intelligent
  surface-aided wireless communications: Adaptive beamforming and experimental
  validations,'' \emph{IEEE Access}, vol.~9, pp. 147\,442--147\,457, 2021.

\bibitem{chen2022calibrated}
H.~Chen, W.~Lin, S.~Wang, W.~Che, and Q.~Xue, ``A calibrated over-the-air
  measurement method for error vector magnitude characterization,'' \emph{IEEE
  Transactions on Instrumentation and Measurement}, vol.~71, pp. 1--8, 2022.

\bibitem{lodro2020near}
M.~Lodro, C.~Smart, G.~Gradoni, A.~Vukovic, D.~Thomas, and S.~Greedy,
  ``Near-field ber and evm measurement at 5.8 ghz in mode-stirred metal
  enclosure,'' \emph{The Applied Computational Electromagnetics Society Journal
  (ACES)}, pp. 1080--1088, 2020.

\bibitem{lodro2021reconfigurable}
M.~Lodro, G.~Gradoni, J.-B. Gros, S.~Greedy, and G.~Lerosey, ``Reconfigurable
  intelligent surface-assisted bluetooth low energy link in metal enclosure,''
  \emph{Frontiers in Communications and Networks}, p.~44, 2021.

\bibitem{gros2020tuning}
J.-B. Gros, P.~del Hougne, and G.~Lerosey, ``Tuning a regular cavity to wave
  chaos with metasurface-reconfigurable walls,'' \emph{Physical Review A}, vol.
  101, no.~6, p. 061801, 2020.

\end{thebibliography}


% \begin{IEEEbiography}{Mir Lodro}
% \end{IEEEbiography}

% \begin{IEEEbiographynophoto}{Gabriele Gradoni}

% % insert where needed to balance the two columns on the last page with
% %\newpage



%\vfill


\end{document}